\documentclass[reprint,amssymb,superscriptaddress, nofootinbib]{revtex4-1}
\usepackage[pdftex]{graphicx}
\usepackage{subfig}
\usepackage{amsmath,amsfonts,amssymb,mathtools}
\usepackage{tasks}
\usepackage{diagbox}
\usepackage{float}
\usepackage{csquotes}
\usepackage{float}
\usepackage{bbm,bm}
\usepackage[normalem]{ulem}
\usepackage{comment}
\usepackage{physics}
\usepackage{xcolor}
\usepackage[utf8]{inputenc}
\usepackage{pgfplots}
\definecolor{blueprl}{RGB}{46,48,146}
\usepgfplotslibrary{groupplots}

\usepackage{lipsum}

\def\ANU{Centre for Quantum Computation and Communication Technology, Department of Quantum Science, Australian National University, Canberra, ACT 2601, Australia.}

  \def\Jena{Institute of Applied Physics, Abbe Center of Photonics, Friedrich-Schiller-Universität Jena, 07745 Jena, Germany}
  
  \def\Fraun{Fraunhofer-Institute for Applied Optics and Precision Engineering IOF, 07745 Jena, Germany}
  \def\MP{Max Planck School of Photonics, 07745 Jena, Germany}
  \def\Astar{Institute of Materials Research and Engineering, Agency for Science Technology and Research (A*STAR), 2 Fusionopolis Way, 08-03 Innovis 138634, Singapore}
  \def\OzGov{Sensors and Effectors Division of Defence Science and Technology Group, Edinburgh SA 5111, Australia.}
\begin{document}
\title{Testing the postulates of quantum mechanics with coherent states of light and homodyne detection}
\author{Lorc{\'a}n O. Conlon}
\email{lorcanconlon@gmail.com}
\affiliation{\ANU}
\author{Angus Walsh}
\affiliation{\ANU}
\author{Yuhan Hua}
\affiliation{\ANU}
\author{Oliver Thearle}
\affiliation{\OzGov}
\author{Tobias Vogl}
\affiliation{\Jena}
\affiliation{\Fraun}
\author{Falk Eilenberger}
\affiliation{\Jena}
\affiliation{\Fraun}
\affiliation{\MP}
\author{Ping Koy Lam}
\affiliation{\ANU}
\affiliation{\Astar}
\author{Syed M. Assad}
\email{cqtsma@gmail.com}
\affiliation{\ANU}

\begin{abstract}
Quantum mechanics has withstood every experimental test thus far. However, it relies on ad-hoc postulates which require experimental verification. Over the past decade there has been a great deal of research testing these postulates, with numerous tests of Born's rule for determining probabilities and the complex nature of the Hilbert space being carried out. Although these tests are yet to reveal any significant deviation from textbook quantum theory, it remains important to conduct such tests in different configurations and using different quantum states. Here we perform the first such test using coherent states of light in a three-arm interferometer combined with homodyne detection. Our proposed configuration requires additional assumptions, but allows us to use quantum states which exist in a larger Hilbert space compared to previous tests. For testing Born's rule, we find that the third order interference is bounded to be $\kappa=0.002\pm0.004$ and for testing whether quantum mechanics is complex or not we find a Peres parameter of $F = 1.0000 \pm 0.0003$ ($F=1$ corresponds to the expected complex quantum mechanics). We also use our experiment to test Glauber's theory of optical coherence.

\end{abstract}
\maketitle

\section{Introduction}
It is well known that quantum mechanics is incompatible with general relativity~\cite{hossenfelder2010experimental}. One proposed solution to this problem is that massive quantum superpositions will decay at a rate proportional to the mass of the system~\cite{carlesso2022present}. This has been experimentally tested, with no evidence for gravitationally induced collapse found so far~\cite{donadi2021underground}. Event formalism was proposed as a description of quantum optical fields in curved space time~\cite{ralph2009quantum}, however, this too has been experimentally refuted~\cite{xu2019satellite}. Alternative theories propose that either quantum mechanics or general relativity is an approximate theory, leading to tests of general relativity where quantum effects are important~\cite{colella1975observation,rosi2017quantum} and tests of quantum mechanics in alternative reference frames~\cite{fink2017experimental,restuccia2019photon}. So far none of these tests have found any clues as to what might be incomplete with either quantum mechanics or general relativity.

\begin{figure}[t]
\includegraphics[width=0.5\textwidth]{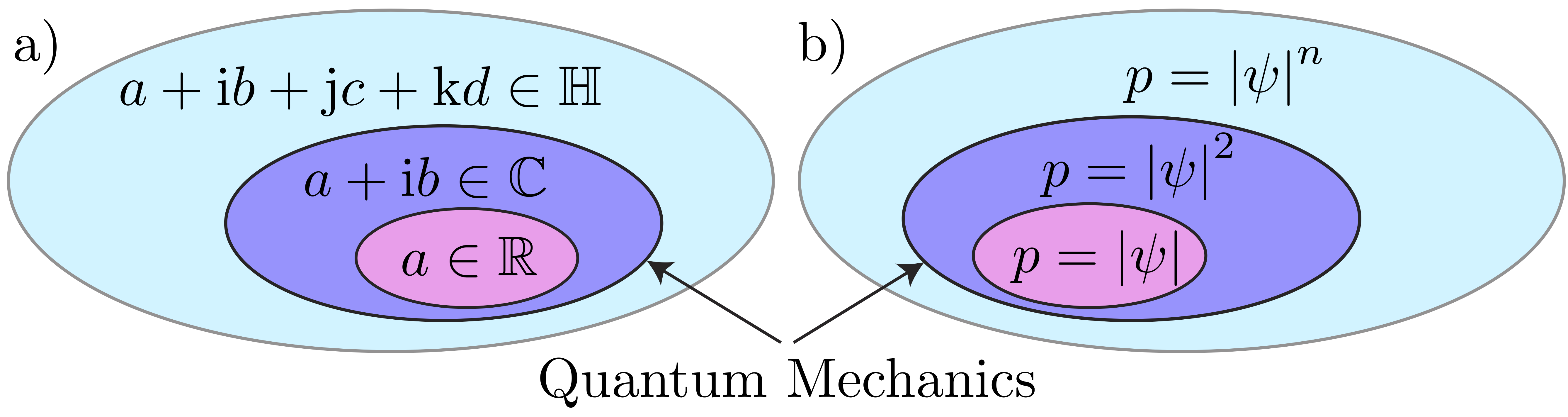}
\caption{\textbf{Ad hoc postulates of quantum mechanics.} a) Quantum states existing in a complex Hilbert space is an \mbox{ad hoc} postulate, as there is no a priori reason to assume that quantum mechanics is complex. For example quantum mechanics could have been described by real numbers or quaternions. b) Another ad hoc postulate is the Born's rule postulate for determining the probabilities of measurement outcomes. Probabilities in quantum mechanics are proportional to the square of the wave function rather than some other power.}
\label{fig:overview}
\end{figure}

Another line of attack is to test the postulates of quantum mechanics. The reasoning behind this is that some of the postulates of quantum mechanics are rather ad hoc, choosing one particular theory out of many possibilities, as shown in Fig.~\ref{fig:overview}. For instance, Born's rule, used to determine the probability of a measurement outcome in quantum mechanics~\cite{born1926quantenmechanik}, is just one possible probability rule out of a family of possible rules~\cite{sorkin1994quantum}. While there have been several attempts to derive Born's rule~\cite{gleason1975measures,cooke1985elementary,zurek2003environment,saunders2004derivation,van2008quantum,masanes2019measurement}, these derivations ultimately require some other underlying assumptions about quantum theory. It is therefore still important to experimentally test Born's rule. This was done in 2010 by Sinha \textit{et al}, who performed the first test explicitly searching for third order interference in quantum mechanics~\cite{sinha2010ruling}. This has led to numerous further experimental tests along this line~\cite{hickmann2011born,sollner2012testing,park2012three,magana2016exotic,kauten2017obtaining,cotter2017search,jin2017experimental,rengaraj2018measuring,barnea2018matter,vogl2021sensitive,pleinert2021testing,chen2021generalized,namdar2023experimental,ahmadi2023quick}. No significant violation of Born's rule has been found so far. 

Another testable postulate of quantum mechanics is that quantum states exist in a complex Hilbert space. For example, one alternative is a quaternonic version of quantum mechanics~\cite{adler1995quaternionic} (quaternions are an extension of complex numbers with different properties -- more detail is provided later in the manuscript). Two possible experiments to distinguish between complex and quaternonic quantum mechanics, based on their different commutation rules, were proposed by Peres in 1979~\cite{peres1979proposed}. One of the proposed tests was carried out using neutrons four years later~\cite{kaiser1984neutron}, where no result was obtained that falsified complex quantum mechanics. More recently, this test was carried out using single photons~\cite{procopio2017single} and once again the complexity of quantum mechanics was verified. Peres' second proposed test has recently been mapped to an experiment based on a three-arm optical interferometer and tested~\cite{gstir2021towards}. Although this particular experiment produced a value which deviates from the expected value given by complex quantum mechanics, multiple plausible reasons were presented for this deviation. Another experiment based on Peres' second proposal was then carried out on a superconducting quantum computer~\cite{sadana2022testing}. Once again, a deviation from the expected result was observed, however, in this case the deviation is explained by the noise of the quantum computer. Thus, to date, Peres' second proposal has not been used to accurately verify that quantum states exist in a complex Hilbert space.


Here we implement both a test of Born's rule and Peres' second proposal using two key tools of Gaussian quantum optics: coherent states and homodyne detection. Under additional assumptions, not required in previous tests, we confirm both Born's rule and that quantum mechanics is indeed complex to a high degree of precision. Our approach contrasts with most previous tests of the postulates of quantum mechanics, which predominantly use single photon sources and detectors. One possible reason for this is the idea that an ideal single photon source can avoid one of the main issues suffered by coherent states: quantum fluctuations or shot noise~\cite{vogl2021sensitive}. However, as we prove, this idea is flawed. We show that the measurement variance for the two differs only slightly. Furthermore, coherent states and homodyne detection offer great prospects for future improved tests of quantum mechanics. For example, homodyne detection can achieve quantum-limited performance at very high frequencies~\cite{rusca2020fast}, which may help future tests avoid low frequency noise. Additionally, as we show, our approach can also be used to test Glauber's theory of coherence, which describes, among other things, the evolution of coherent states~\cite{glauber1963photon,glauber1963quantum,glauber1963coherent}. Finally, coherent states and homodyne detection are readily available and easily implemented, leading to their application in various protocols~\cite{furusawa1998unconditional,haw2016surpassing,zhao2020high}. These advantages, combined with the very precise values we obtain, confirming that quantum mechanics behaves as expected, serve to validate the role that Gaussian quantum optics has to play in present and future tests of quantum mechanics.

\section{Experimental Proposals}
\label{sec:expprop}
Conventional experiments for testing Born's rule and the complexity postulate of quantum mechanics rely on inferring the probability of a detector click in a certain experimental configuration. For tests using single photon states, these probabilities are often related to the photon rate, the number of photon clicks per second. By normalising the photon rate to the total number of input photons, the probability of a detector click in a single experimental run can be obtained. When using homodyne detection, we do not have direct access to the photon counting statistics. Instead, we infer the probabilities of interest by measuring the mean photon number in a given configuration and normalising to the mean photon number of the input state. In this section we shall first describe in detail how we obtain the probabilities of interest using homodyne detection, before describing the various tests we can perform with our set-up. Finally, we shall compute the measurement variance of our proposal and compare this to the measurement variance of conventional proposals using single photon states and detectors.

\subsection{Coherent states and homodyne detection}
\label{sec:CShomo}
\begin{figure*}[t]
\includegraphics[width=0.9\textwidth]{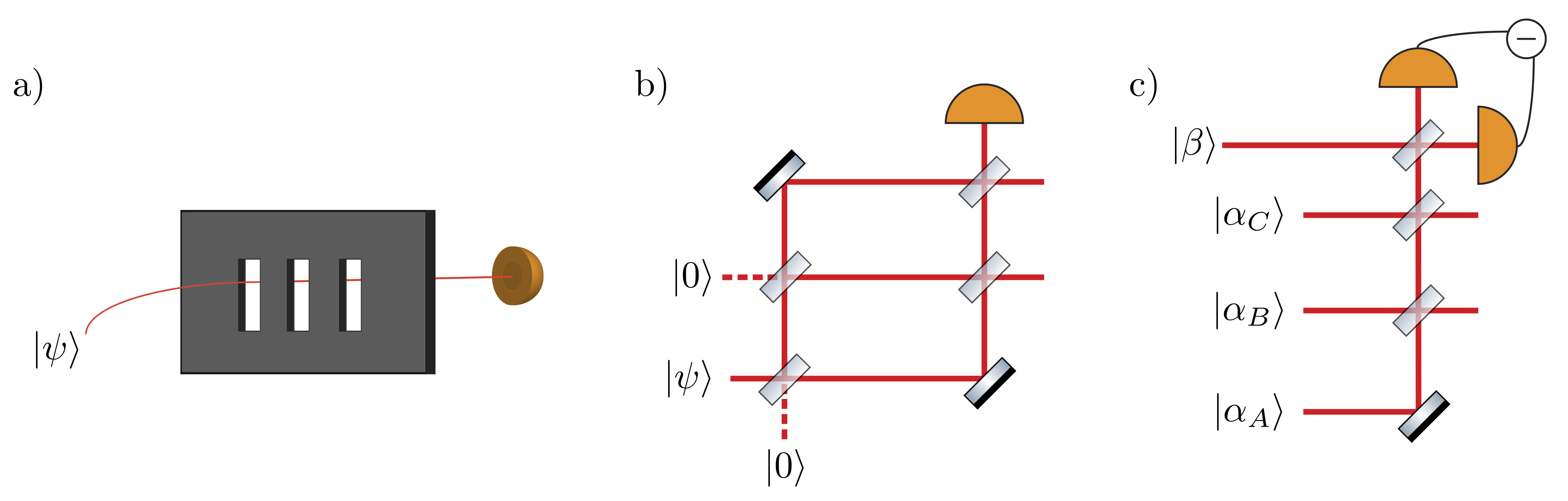}
\caption{\textbf{Three interferometer types for testing the postulates of quantum mechanics.} a) Most tests in recent years have used physical slits, which can be blocked to give the different path configurations. b) Some tests have used an optical interferometer with a single input state, and the three optical paths can be blocked. The output states are measured with a single photon detector. c) Our test uses an alternative configuration. We use three input coherent states in an open interferometer. Instead of blocking and opening the different paths in the interferometer we turn off and on the coherent states. The output state is measured with homodyne detection.}
\label{fig:interf}
\end{figure*}

Coherent states are one example of a quantum state from the family of continuous variable Gaussian states. Such states can be defined by their first and second moments of the amplitude and phase quadratures of the electromagnetic field, denoted $\hat{x}$ and $\hat{p}$ respectively. The quadratures are defined in terms of the creation and annihilation operators, which we denote $\hat{a}$ and $\hat{a}^\dagger$ respectively, as $\hat{x}=\hat{a}+\hat{a}^\dagger$ and $\hat{p}=-\mathrm{i}(\hat{a}-\hat{a}^\dagger)$. Coherent states are minimum uncertainty states, in that, $\Delta\hat{x}^2=\Delta\hat{p}^2=1$, where $\Delta\hat{x}^2$ is the variance in the $\hat{x}$ quadrature and similarly for the $\hat{p}$ quadrature.

Most experiments aimed at testing the postulates of quantum mechanics in recent years have focused on interferometers of the types shown in Fig.~\ref{fig:interf} a) and b). For our experiment, we will consider the interferometer shown in Fig.~\ref{fig:interf} c). There are two major differences between our interferometer and the interferometers shown in Fig.~\ref{fig:interf} a) and b). The first is that we use three input states, $\ket{\alpha_A}$, $\ket{\alpha_B}$ and $\ket{\alpha_C}$, which can be turned on and off instead of a single input state with paths which can be blocked. This can be justified as if we replace $\ket{\psi}$ in Fig.~\ref{fig:interf} b) with a coherent state, the quantum state in each of the three arms will be a coherent state~\cite{weedbrook2012gaussian}. The output state of our interferometer will depend on which coherent states are turned on and which are turned off. We use $c_{i}=0,1$ to indicate whether the coherent state on path $i$ is turned on ($c_i=1$) or turned off ($c_i=0$). The output state also depends on the beamsplitters used to mix the three coherent states. We use beamsplitters with transmissivities of $1/\sqrt{2}$ and $\sqrt{2/3}$ to ensure that the output coherent state is given by
\begin{equation}
\label{eq:cohout}
\ket{\alpha_\text{out}}=\ket{\frac{1}{\sqrt{3}}(c_A\alpha_A+c_B\alpha_B+c_C\alpha_C)}\;.
\end{equation}
Thus the interferometers in in Fig.~\ref{fig:interf} b) and c) are equivalent for the coherent states we are considering. However, it should be noted that this equivalence is not necessary for the tests we are considering.

The second difference is in the detection stage. We propose to use homodyne detection instead of an optical power measurement or photon counting. Homodyne detection is designed to measure either of the quadratures, $\hat{x}$ or $\hat{p}$, of a given input state. We write the state to be measured as $\ket{\alpha_\text{out}}=\ket{\alpha_\mathrm{r}+\mathrm{i}\alpha_\mathrm{i}}$, where $\alpha_\mathrm{r}$ and $\alpha_\mathrm{i}$ are real numbers. Note that, for now, we are assuming standard, complex quantum mechanics to describe the theoretical background. The state to be measured is mixed with a much brighter coherent state with a controllable phase, denoted $\ket{\text{e}^{\mathrm{i}\phi}\beta}$, on a beamsplitter of transmissivity $t$. After mixing this state on a beamsplitter with the bright local oscillator, $\ket{\text{e}^{\mathrm{i}\phi}\beta}$, two states are produced
\begin{equation}
\ket{t\alpha_\text{out}+\sqrt{1-t^2}\text{e}^{\mathrm{i}\phi}\beta}\;,
\end{equation}
and
\begin{equation}
\ket{-\sqrt{1-t^2}\alpha_\text{out}+t\text{e}^{\mathrm{i}\phi}\beta}\;.
\end{equation}
Both of these states are detected on a photodetector which is implementing the number operator $\hat{n}=\hat{a}^\dagger\hat{a}$. The photocurrent produced by each detector, $I_i$, is proportional to the gain of each detector, $g_i$ and the expectation value of the number operator, $\langle \hat{n}\rangle$, i.e. $I_i\propto g_i\langle \hat{n}\rangle$. In ideal homodyne detection $g_1=g_2$ and $t=1/\sqrt{2}$, which gives the subtracted photo current as
\begin{equation}
\label{eq:homoout}
I_1-I_2\propto \beta(\alpha_\mathrm{r}\text{cos}(\phi)+\alpha_\mathrm{i}\text{sin}(\phi))\;.
\end{equation}
The full calculation of the subtracted photocurrent is provided in Appendix~\ref{apen:homo}. Strictly speaking, Eq.~\eqref{eq:homoout} is true if we measure the DC power directly, whereas in our experiments we create and measure our quantum states at high frequencies. However, Eq.~\eqref{eq:homoout} is still valid at higher frequencies, to within the linearisation approximation used in quantum optics. By choosing $\phi=0$ or $\phi=\pi/2$ we can measure the $\hat{x}$ and $\hat{p}$ quadratures respectively. For a coherent state, $\ket{\alpha},$ the mean photon number is given by $\langle \hat{n}\rangle=\abs{\alpha}^2$. Hence, if we only generate displacements in one quadrature, i.e. we set either $\alpha_\mathrm{r}=0$ or $\alpha_\mathrm{i}=0$, we can infer the mean photon number from the square of the homodyne measurement results. The mean photon number, or photon rate, is proportional to the probability we wish to measure. In this way we can infer the probabilities for each of the experimental configurations used to test quantum mechanics. \textbf{We note that this and the linearisation approximation mentioned above are essential additional assumptions of our test, which are not required for previous tests.} 


%

\subsection{Sorkin test}
\label{subsecsorkin}
The original test of Born's rule was based on a triple slit experiment, of the type shown in Fig.~\ref{fig:interf} a). There are three possible paths which the quantum state can take to reach the detector, all of which can be opened and closed. This can be mapped to an optical interferometer as shown in Fig.~\ref{fig:interf} b), which as mentioned in the previous section, is equivalent to the interferometer shown in Fig.~\ref{fig:interf} c) for coherent state inputs. Blocking the different possible paths in Figs.~\ref{fig:interf} a) and b) is equivalent to turning on and off the coherent states in Fig.~\ref{fig:interf} c), which gives rise to the output state in Eq.~\eqref{eq:cohout}. For our experiment, we generate our coherent states only in the $\hat{p}$ quadrature, i.e. $\alpha_A$, $\alpha_B$ and $\alpha_C$ are all imaginary numbers. The mean photon number of this state is therefore given by the square of the measurement outcomes along the $\hat{p}$ quadrature
\begin{equation}
\langle n\rangle = \frac{1}{3}\abs{c_A\alpha_A+c_B\alpha_B+c_C\alpha_C}^2\;.
\end{equation}
We shall denote the mean photon number of the input state as $\langle n_\text{T}\rangle$ and the mean photon number of the output state with the $i$th coherent states turned on as $\langle n_i\rangle$, i.e. $\langle n_A\rangle=\abs{\alpha_A}^2/3$. Therefore, the probability of obtaining an output photon per input photon with the $i$th coherent states turned on is given by $P_i=\langle n_i\rangle/\langle n_\text{T}\rangle$. For our proposal, we can choose $\alpha_A=\alpha_B=\alpha_C=\mathrm{i}/\sqrt{3}$, so that $P_A=P_B=P_C=1/9$, $P_{A,B}=P_{A,C}=P_{B,C}=4/9$ and $P_{A,B,C}=1$.

One of the most surprising features of quantum mechanics is revealed in a double slit experiment, the probability with two paths open is not equal to the sum of the probabilities of the individual paths. This is revealed in the fact that second-order interferences are non-zero in quantum mechanics
\begin{equation}
I_{A,B}=P_{A,B}-P_{A}-P_B\neq0\;.
\end{equation} 
Intriguingly however, third-order interferences are zero in quantum mechanics
\begin{equation}
I_{A,B,C}=P_{A,B,C}-P_{A,B}-P_{A,C}-P_{B,C}+P_{A}+P_B+P_C=0\;.
\end{equation} 
This is a consequence of Born's rule. By measuring the probabilities in the above equation, we can therefore experimentally determine whether $I_{A,B,C}=0$ holds in practice. To do so, we define
\begin{equation}
\epsilon=\langle n_{A,B,C}\rangle-\langle n_{A,B}\rangle-\langle n_{A,C}\rangle-\langle n_{B,C}\rangle+\langle n_{A}\rangle+\langle n_{B}\rangle+\langle n_{C}\rangle\;,
\end{equation}
which represents the third order interference and
\begin{equation}
\begin{split}
\delta=&\abs{\langle n_{A,B}\rangle-\langle n_{A}\rangle-\langle n_{B}\rangle}+\abs{\langle n_{A,C}\rangle-\langle n_{A}\rangle-\langle n_{C}\rangle}\\&+
\abs{\langle n_{B,C}\rangle-\langle n_{B}\rangle-\langle n_{C}\rangle}\;,
\end{split}
\end{equation}
which represents the sum of the second order interference terms. This allows us to define the following quantity
\begin{equation}
\kappa=\frac{\epsilon}{\delta}\;,
\end{equation}
which represents the normalised third-order interference.

\subsection{Peres' proposal}
To understand Peres' proposal, and how it distinguishes complex numbers and quaternions, it is necessary to first describe some of the basic properties of quaternions. Quaternions are an extension of complex numbers. Whereas a complex number can be described by two real numbers, $a$ and $b$, as $a+\mathrm{i}b$, a quaternion is described by four real numbers, $a$, $b$, $c$, and $d$, as~\cite{hamilton1848xi}
\begin{equation}
a+\mathrm{i}b+\mathrm{j}c+\mathrm{k}d\;.
\end{equation}
Quaternions satisfy 
\begin{equation}
\mathrm{i}^2=\mathrm{j}^2=\mathrm{k}^2=\mathrm{i}\mathrm{j}\mathrm{k}=-1\;.
\end{equation}
The key difference between complex numbers and quaternions is that complex numbers commute whereas quaternions do not commute in general. This property ensures that there exist testable differences between the predictions of quantum mechanics when described using complex numbers and quaternions. 

\begin{figure*}[t]
\includegraphics[width=0.7\textwidth]{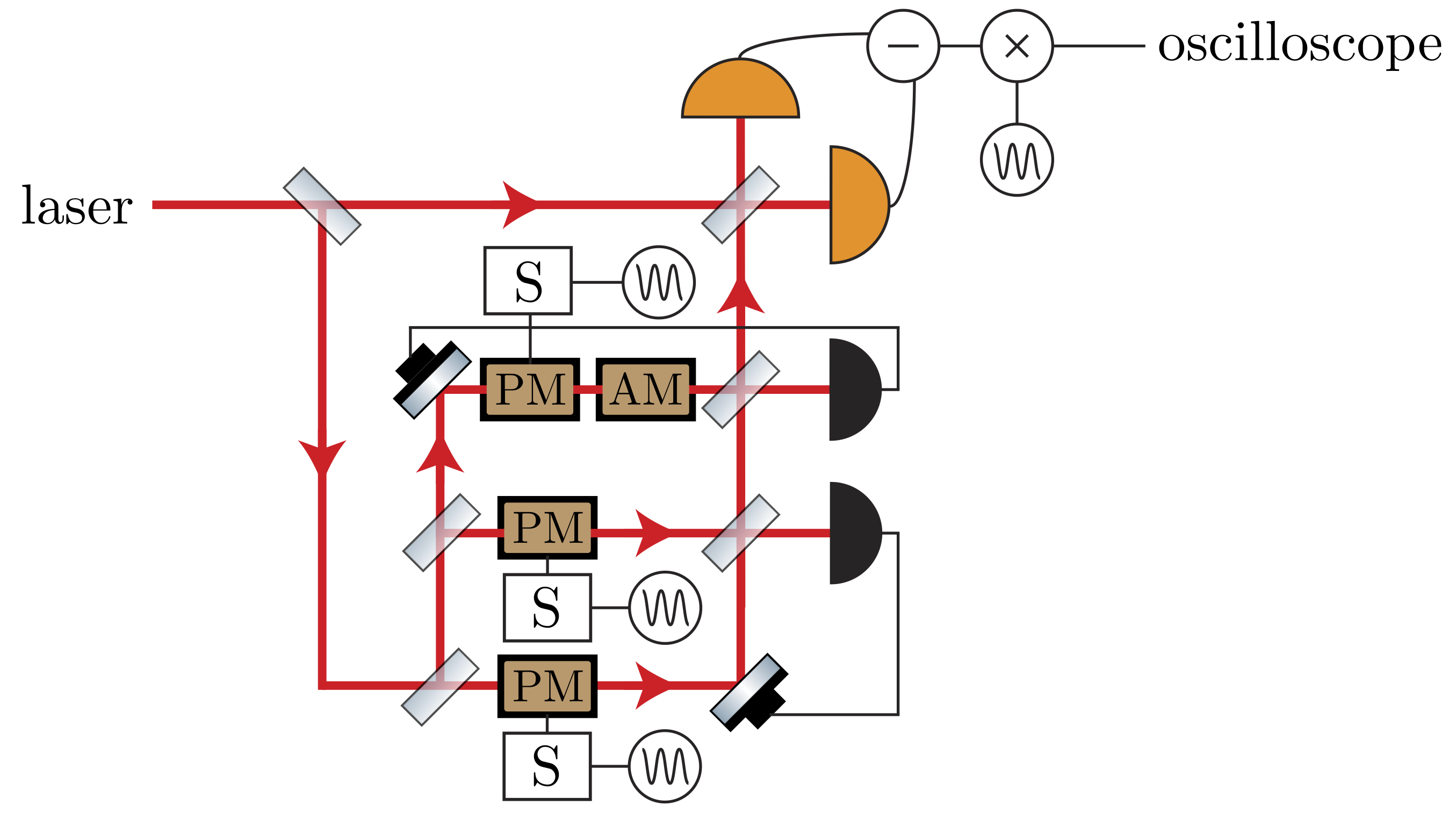}
\caption{\textbf{Experimental set-up.} The laser is split into four paths, one for the local oscillator for homodyne detection and three for the interferometer arms. Coherent states are created by a phase modulator (PM) with a 4 MHz sine wave. These states are turned on and off using an RF switch (S). The black detectors on the otherwise unused ports of the interferometer are used for locking the relative phase of the three arms of the interferometer. This is done through feedback to piezoelectric mirrors. The orange detectors are used for homodyne detection. After detection the photocurrents are subtracted and mixed down with a 4 MHz sine wave to produce the output signal, which is sent to an oscilloscope to be recorded.}
\label{fig:expSU}
\end{figure*}

To examine these differences, we now describe the version of Peres' proposal presented in Ref.~\cite{gstir2021towards}. We consider a three-arm interferometer, with paths A, B and C providing phase shifts of $\phi_A$, $\phi_B$ and $\phi_C$ respectively. Defining $\Delta\phi_{i,j}$ as $\Delta\phi_{i,j}=\phi_i-\phi_j$, we have
\begin{equation}
\Delta\phi_{A,B}+\Delta\phi_{B,C}+\Delta\phi_{C,A}=0\;.
\end{equation}
Using this relation, and some trigonometry identities we arrive at
\begin{equation}
\label{eq:F}
F\equiv \beta^2+\gamma^2+\zeta^2-2\beta\gamma\zeta=1\;,
\end{equation}
where $\beta=\text{cos}(\Delta\phi_{B,C})$, $\gamma=\text{cos}(\Delta\phi_{C,A})$ and $\zeta=\text{cos}(\Delta\phi_{A,B})$\footnote{Here we use $\beta$, $\gamma$ and $\zeta$ in place of the more commonly used $\alpha$, $\beta$ and $\gamma$ to reserve $\alpha$ for describing coherent states.}. The above relation holds for regular complex quantum mechanics. However, if we allow phase shifts to be described by quaternions, we can have $F<1$. A simple example using quaternion phase shifts which gives $F=0$ is presented in Appendix~\ref{apen:Feg}.

Therefore, by experimentally determining the quantities $\beta$, $\gamma$ and $\zeta$, we can test for hypercomplex quantum mechanics. These quantities can be determined from the probabilities described in the previous section. For a two-path interferometer, the probability of detecting a photon is given by 
\begin{equation}
P_{A,B}=P_A+P_B+2\sqrt{P_AP_B}\zeta\;.
\end{equation}
As $P_{A,B}$, $P_{A}$ and $P_{B}$ are experimentally accessible (through $\langle n_{A,B}\rangle$, $\langle n_{A}\rangle$ and $\langle n_{B}\rangle$), this gives us a way to determine $\zeta$ experimentally. As $\beta$ and $\gamma$ can be determined in a similar manner, this gives us a simple way to experimentally determine $F$ to verify if Eq.~\eqref{eq:F} is satisfied.

It is important to note that because we are not accessing the mean photon number directly, our test of hypercomplex quantum mechanics requires an additional assumption. We need to assume that if coherent states have some quaternion component, i.e. we extend $\ket{\alpha_\mathrm{r}+\mathrm{i}\alpha_\mathrm{i}}$ to $\ket{\alpha_\mathrm{r}+\mathrm{i}\alpha_\mathrm{i}+\mathrm{j}\alpha_\mathrm{j}+\mathrm{k}\alpha_\mathrm{k}}$, then if we measure the $\hat{p}$ quadrature we will find that $\abs{I_1-I_2}\propto\abs{\mathrm{i}\alpha_\mathrm{i}}$ is extended to $\abs{I_1-I_2}\propto\abs{\mathrm{i}\alpha_\mathrm{i}+\mathrm{j}\alpha_\mathrm{j}+\mathrm{k}\alpha_\mathrm{k}}$. In Appendix~\ref{apen:homocompl} we provide a heuristic justification for why this assumption is reasonable when testing for hypercomplex quantum mechanics.


\subsection{Test of Glauber's theory of coherence}
As we are using coherent states and homodyne detection to test the postulates of quantum mechanics, we can also test Glauber's theory of coherence~\cite{glauber1963photon,glauber1963quantum,glauber1963coherent}. According to textbook quantum optics, the output coherent state from our interferometer is given by Eq.~\eqref{eq:cohout}. For $\alpha_A=\alpha_B=\alpha_C=\mathrm{i}/\sqrt{3}$, the expectation value of a measurement of the $\hat{p}$ quadrature is
\begin{equation}
\langle \hat{p}\rangle\propto\frac{-\mathrm{i}}{\sqrt{3}}(c_A\alpha_A+c_B\alpha_B+c_C\alpha_C)\;.
\end{equation}
Defining $\langle \hat{p}_i\rangle$ as the $\hat{p}$ quadrature measurement with the $i$th coherent state turned on, we see that the following normalised quantities should be zero 
\begin{equation}
\begin{split}
G_{A,B}&=\frac{\langle \hat{p}_{A,B}\rangle-\langle \hat{p}_{A}\rangle-\langle \hat{p}_{B}\rangle}{\abs{\langle \hat{p}_{A}\rangle}+\abs{\langle \hat{p}_{B}\rangle}}\;,\\
G_{A,C}&=\frac{\langle \hat{p}_{A,C}\rangle-\langle \hat{p}_{A}\rangle-\langle \hat{p}_{C}\rangle}{\abs{\langle \hat{p}_{A}\rangle}+\abs{\langle \hat{p}_{C}\rangle}}\;,\\
G_{B,C}&=\frac{\langle \hat{p}_{B,C}\rangle-\langle \hat{p}_{B}\rangle-\langle \hat{p}_{C}\rangle}{\abs{\langle \hat{p}_{B}\rangle}+\abs{\langle \hat{p}_{B}\rangle}}\;,\\
G_{A,B,C}&=\frac{\langle \hat{p}_{A,B,C}\rangle-\sum_{i,j\in A,B,C, i\neq j}\langle \hat{p}_{i,j}\rangle+\sum_{i\in A,B,C}\langle \hat{p}_{i}\rangle}{\abs{\langle \hat{p}_{A}\rangle}+\abs{\langle \hat{p}_{B}\rangle}+\abs{\langle \hat{p}_{C}\rangle}}\;.
\end{split}
\end{equation}
As these quantities are all directly accessible, we can perform a test of Glauber's theory of coherence.

\subsection{Quantum noise comparison}
Quantum metrology aims to use quantum resources to perform sensing or estimation tasks with a better precision than is possible using only classical resources~\cite{marciniak2022optimal,conlon2023approaching}. For example, single photon states can significantly outperform coherent states for some sensing tasks~\cite{sabines2019twin}. Thus, it seems natural to assume that for tests of the postulates of quantum mechanics, an ideal single photon source does not suffer from shot noise, which is one of the main issues that affects coherent states~\cite{vogl2021sensitive}. However, by calculating the measurement variance of our proposed experiment and an experiment using an ideal single photon source, we now show that an ideal single photon source will suffer from shot noise in such tests\footnote{Here we use shot noise to refer to all noise which stems from the inherent probabilistic nature of quantum mechanics.}.

\begin{figure*}[t]
\includegraphics[width=0.95\textwidth]{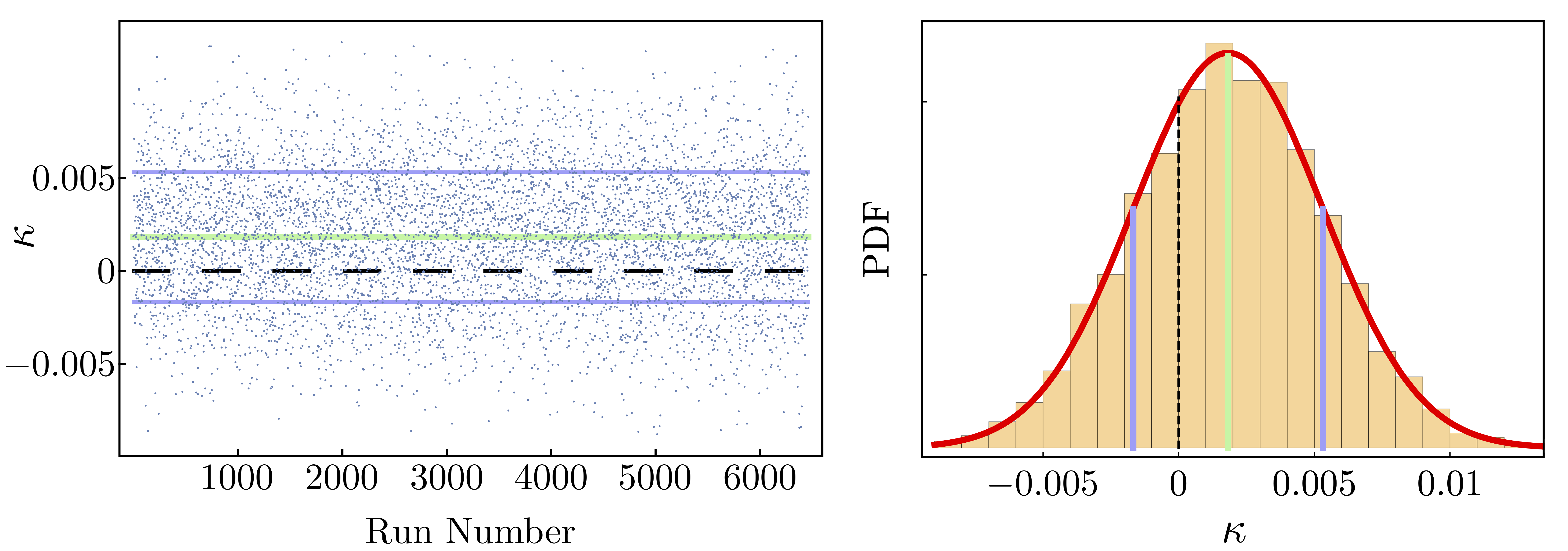}
\caption{\textbf{Experimental results for testing Born's rule.} The figure on the left shows the results of 6462 individual experiments. The dashed black line shows the expected value of $\kappa=0$ predicted by quantum mechanics. The green line shows the mean of the experimental data and the blue lines show one standard deviation. The figure on the right is the histogram of the data in the left figure. The red line shows a normal distribution fitted to the data. Before filtering there were 6517 data points.}
\label{fig:data1}
\end{figure*}

We compare the set-up shown in Fig.~\ref{fig:interf} b) using a single photon state, $\ket{\psi}=\ket{1}$, with the set-up shown in Fig.~\ref{fig:interf} c). To make a fair comparison, in terms of the resources used, we set $\alpha_A=\alpha_B=\alpha_C=1/\sqrt{3}$. This ensures that for both set-ups the mean number of photons used in the test are equal. We choose the beamsplitter transmissivities on the left hand side of the interferometer in Fig.~\ref{fig:interf} b) such that the state inside the interferometer is
\begin{equation}
\ket{\psi_{\text{inside}}}=\frac{1}{\sqrt{3}}(\ket{100}+\ket{010}+\ket{001})\;,
\end{equation}
where $\ket{n_An_Bn_C}$ represents the state with $n_i$ photons in path $i$. The beamsplitters on the right hand side of the interferometer can then be chosen (assuming standard quantum mechanics) such that $P_A=P_B=P_C=1/9$, $P_{A,B}=P_{B,C}=P_{A,C}=4/9$ and $P_{A,B,C}=1$. There are seven probabilities to measure in the test of Born's rule and six to measure in the test of hypercomplex quantum mechanics. We assume that we have $N$ photons available to determine each probability. For any given configuration of open and closed paths, the probability can be estimated from the number of photon clicks observed, $n_i$, as
\begin{equation}
\tilde{P}_i=\frac{n_i}{N}\;.
\end{equation}
The number of clicks observed can be modelled as a binomial distribution. The variance in the estimate of each probability is therefore given by
\begin{equation}
\sigma_{P_i}^2=\frac{P_i(1-P_i)}{N}\;.
\end{equation} 
Using error propagation formula it is then possible to obtain the error in our estimates of $\epsilon$, $\delta$, $\beta$, $\gamma$ and $\zeta$, and by extension $\kappa$ and $F$. However, it is not necessary to explicitly do so as a reduced error in estimating the individual probabilities translates into a reduced error in estimating either $\kappa$ or $F$. Hence, rather than presenting an unwieldy formula for the error in estimating either of these quantities, we shall simply make a direct comparison of the error in estimating the individual probabilities.

For estimating each probability with coherent states, we are interested in inferring the mean photon number. We do this by first performing homodyne detection, which produces a normal distribution with a mean which depends on the experimental configuration and a normalised variance of one. We then take the mean of this distribution, divide by two and square this value which produces an estimate for the mean photon number. The mean photon number estimates thus follow a squared Gaussian distribution (or equivalently, a non-central $\chi^2$ distribution). For a Gaussian distribution with mean $\mu$ and standard deviation $\sigma$, the square of this distribution will have variance
\begin{equation}
\sigma_{\mathcal{N}^2}^2=2\sigma^4+4\sigma^2\mu^2\;.
\label{Eq:BRprobdistsquared}
\end{equation}
As the mean number of photons used in this test is 1, the estimate of the mean photon number in any given configuration is equal to the estimate of the probability. From the expected mean value of the homodyne detection we can therefore determine the variance in estimating the probability. Thus, using coherent states, we have $\sigma_{P_A}^2=\sigma_{P_B}^2=\sigma_{P_C}^2=17/72N$. For comparison, with a single photon state, we have $\sigma_{P_{A}}^2=\sigma_{P_{B}}^2=\sigma_{P_{C}}^2=8/81N$, which differs by a factor $153/64\approx 2.5$. With two paths open, using coherent states we have $\sigma_{P_\text{A,B}}^2=\sigma_{P_{A,C}}^2=\sigma_{P_{B,C}}^2=41/72N$. Using an ideal single photon source, we have $\sigma_{P_{A,B}}^2=\sigma_{P_{A,C}}^2=\sigma_{P_{B,C}}^2=20/81N$ and so the two differ by a factor $369/160\approx2.5$ in this case. Hence, it is clear that both systems will suffer from a very similar amount of shot noise. Ultimately, as quantum noise can be reduced by repeating the experiment, other systematic noise sources are likely to be the dominant noise in such experiments.

%

Here we have only calculated the variance for two fixed input states. However, in future work it would be useful to consider the quantum Fisher information of these states and to find measurements which saturate the quantum Fisher information. It would also be beneficial to know what states are optimal, from a quantum metrology viewpoint, for performing such tests. In Appendix~\ref{apen:squeezing}, we present some initial simulations along this line: we calculate the measurement variance when using a displaced squeezed state as our input, rather than simply a coherent state. We show that, subject to a fixed mean photon number constraint, there is a benefit to using displaced squeezed states. With recent developments in highly stable sources of squeezed light~\cite{shajilal202212}, the advantages of using squeezed light in these tests may soon be realised.

\begin{figure*}[t]
\includegraphics[width=0.95\textwidth]{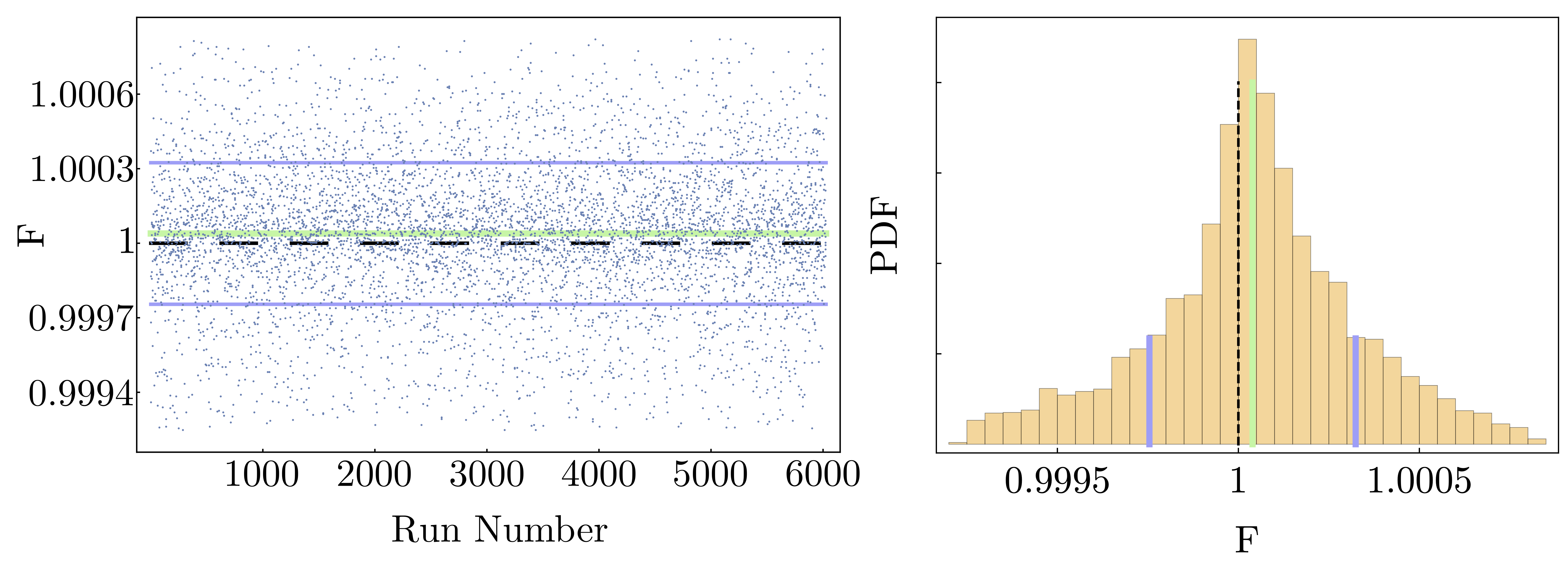}
\caption{\textbf{Experimental data testing whether quantum mechanics is complex.} The figure on the left shows the results of 6027 individual experiments. The dashed black line shows the expected value of $F=1$ predicted by quantum mechanics. The green line shows the mean of the experimental data and the blue lines show one standard deviation. The figure on the right is the histogram of the data in the left figure. Before filtering there were 6517 data points.}
\label{fig:data2}
\end{figure*}

\section{Experimental results}
\label{sec:expresults}
\subsection{Experimental set-up}
A schematic of our experiment is shown in Fig.~\ref{fig:expSU}. We use a 1064 nm laser, which is split into four paths, three for the arms of the interferometer and one for the local oscillator in the homodyne detection stage. Coherent states are generated using electro-optic phase modulators, modulated at 4 MHz with a sine wave. On one arm of the interferometer we also have an amplitude modulator, to ensure we can lock to the $\hat{p}$ quadrature. Additional detectors are placed on two of the output ports of the interferometer to lock the relative phase of the interferometer. The output coherent states are mixed on a 50:50 beamsplitter with the bright local oscillator. The local oscillator power is set to be at least 100 times greater than the signal power. After mixing with the local oscillator, the coherent states are detected with photodetectors and the photocurrents are subtracted from one another. The subtracted photocurrent is then mixed down at 4 MHz to extract the signal which is measured on an oscilloscope. The measurement bandwidth is approximately 100 kHz. RF switches are used to turn on and off the coherent states.

\subsection{Experimental procedure}
We generate coherent states approximately equal to $\ket{\mathrm{i}}$ on each arm of the interferometer. A single run of the experiment consists of switching through all eight possible configurations, which takes approximately ten seconds, with the major bottleneck being the data saving process. For each configuration we save approximately $1.8\times10^6$ data points. After acquiring the data we apply a filter to remove any points where the interferometer came unlocked. With this filtered data, we use two different approaches to determine the necessary probabilities in the tests of Born's rule and hypercomplex quantum mechanics. To remove systematic errors, we can either correct the mean of the homodyne data directly, by subtracting the mean value with all coherent states turned off. Or, we can correct the mean photon number directly, by subtracting the measured mean photon number with all coherent states turned off, i.e. $\langle n_{A}\rangle_\text{corrected}=\langle n_{A}\rangle-\langle n_{0}\rangle$. The final data is then filtered again to remove outliers. Surprisingly, as we shall see, for testing Born's rule the second approach works best, but for testing whether quantum mechanics is complex or not, the first approach works best.

\subsection{Experimental results}
\subsubsection{Born's rule test experimental results} 
The results from our experiment for testing Born's rule are shown in Fig.~\ref{fig:data1}. When we correct the mean photon number directly, we obtain a value of $\kappa=0.002\pm0.004$, where the uncertainties correspond to one standard deviation. Thus our results are consistent with textbook quantum mechanics which only allows for second order interference. The distribution of estimated $\kappa$ values follows a normal distribution as shown in the right hand side of Fig.~\ref{fig:data1}. It is worth noting that if we correct the mean of the homodyne data directly, as opposed to correcting the mean photon number, we obtain a value of $\kappa=0.006\pm 0.004$. 

\begin{figure*}[t]
\includegraphics[width=1\textwidth]{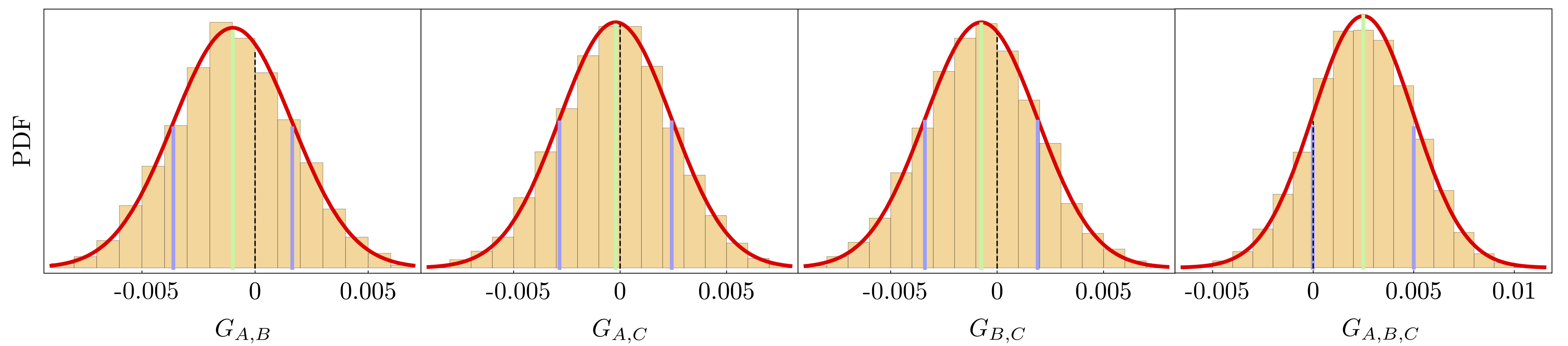}
\caption{\textbf{Experimental data for testing Glauber's theory of coherence.} The four figures correspond to four different tests of Glauber's theory of optical coherence. In all figures the dashed black line shows the expected value of $G=0$, the green line is the mean value and the blue lines correspond to one standard deviation. The red lines show fitted normal distributions. From left to right, the figures correspond to 6464, 6465, 6477 and 6471 data points. Before filtering there were 6517 data points for all figures.}
\label{fig:data3}
\end{figure*}

\subsubsection{Hypercomplex quantum mechanics test results} 
Using the same data set, we can also evaluate the quantity $F$ in Eq.~\eqref{eq:F}, to test whether quantum mechanics is complex. When correcting the mean of the homodyne data (as opposed to correcting the mean photon number as used in the test of Born's rule), we obtain a value of $F = 1.0000 \pm 0.0003$. This data is presented in Fig.~\ref{fig:data2}. As can be seen in the right hand figure, the data in this case does not form a normal distribution. One possible explanation for this is that $F$ is constructed from a different combination of the probabilities compared to $\kappa$.

\subsubsection{Results for testing Glauber's theory of coherence} 

Finally, we use the same data set to test Glauber's theory of coherence in a range of settings. Using two path interference, we find $G_{A,B}=-0.001\pm0.003$, $G_{A,C}=-0.0002\pm0.003$ and $G_{B,C}=-0.0007\pm0.003$. Within the error bars these values are all 0 as expected. However, all of these values are negative, which could hint at some systematic issue that we have not yet resolved. Using three path interference, we find $G_{A,B,C}=0.003\pm0.003$. This data is presented in Fig.~\ref{fig:data3}. As can be seen, in all cases, the data follows a normal distribution.

\subsection{Possible Sources of Error}
\label{sec:error}
Experimental imperfections when using homodyne detection can give apparent non-zero third order interference in tests of Born's rule. This is because imperfections result in the homodyne output deviating from Eq.~\eqref{eq:homoout}, meaning the inferred probability will deviate from the true probability. Note that this is different to previous tests of Born's rule, which have shown several possible scenarios where the third order interference may be non-zero~\cite{magana2016exotic,pleinert2021testing,namdar2023experimental}. These are not violations of Born's rule, as they are exploring alternative physical regimes using multiple particles or non-linear evolution. The potential violations when using imperfect homodyne detection are different and  correspond to incorrectly inferring the probabilities of interest. In this section, we explore the potential magnitude of the apparent Born rule violation caused by imperfect homodyne detection.

\begin{figure*}[t]
\includegraphics[width=0.9\textwidth]{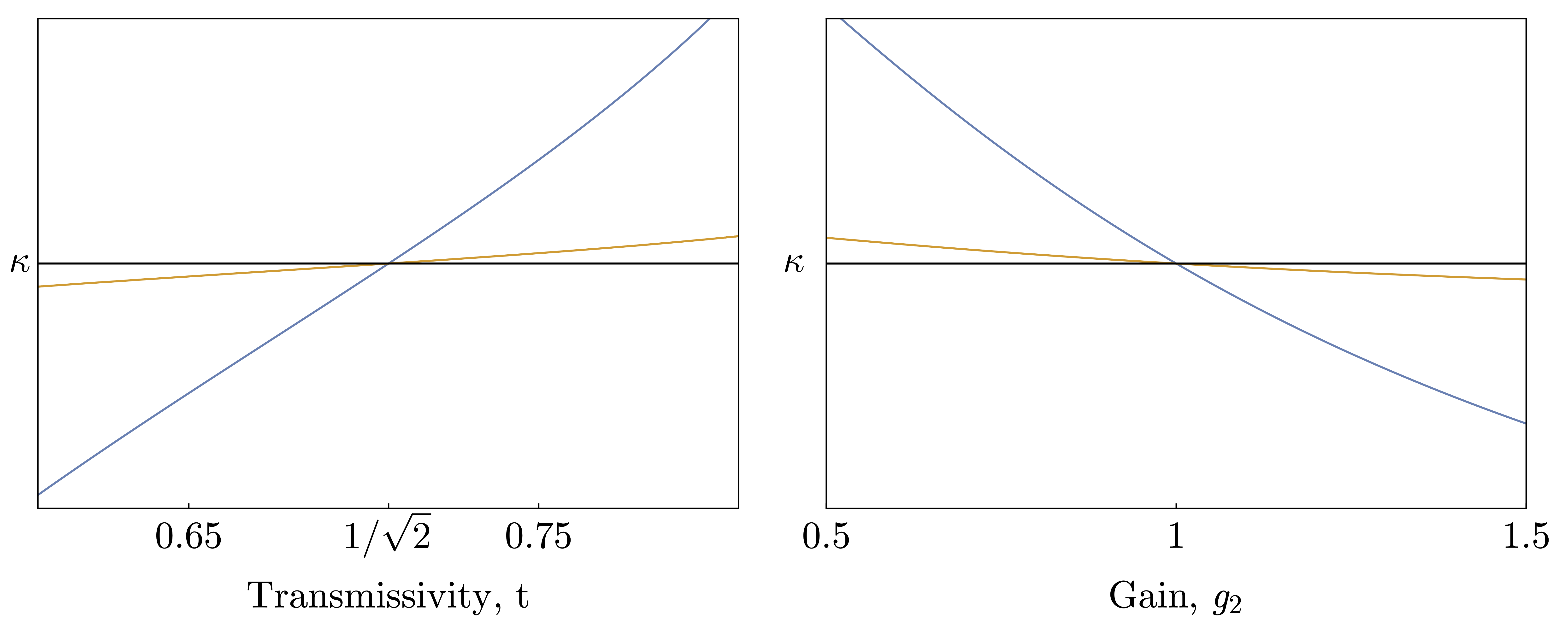}
\caption{\textbf{Possible sources apparent third order interference with imperfect homodyne detection.} Both plots use $\abs{\alpha}=0.1$. The blue and yellow lines correspond to $\abs{\beta}=10^3\abs{\alpha}$ and $\abs{\beta}=10^4\abs{\alpha}$ respectively. The left figure shows the effect of changing the homodyne beamsplitter transmissivity with $g_1=g_2$ and the right figure shows the effect of changing the gain of one detector with $t=1/\sqrt{2}$ and $g_1=1$.}
\label{fig:homoviolation}
\end{figure*}

Equation.~\eqref{eq:realhomo} gives the output of a homodyne measurement in the general scenario, allowing for an imperfect experiment. In the ideal case, with $g_1=g_2$ and $t=1/\sqrt{2}$, the homodyne output reduces to that of Eq.~\eqref{eq:homoout}. In Fig.~\ref{fig:homoviolation} we examine how the value of $\kappa$ changes as we deviate from these ideal parameters. Exact values of $\kappa$ are not shown as the value of $\kappa$ depends on many other parameters such as the gain of both detectors and the local oscillator strength. However, it is clear from Fig.~\ref{fig:homoviolation}, that imperfect homodyne detection can result in apparent violations of Born's rule. Hence, great care is required to avoid this issue.

\section{Conclusion}
\label{sec:conc}
In this paper we have performed multiple tests of quantum mechanics using coherent states and homodyne detection. For testing Born's rule, we have determined a value of $\kappa=0.002\pm0.004$ and for testing whether quantum mechanics is complex or not we have determined a value of $F = 1.0000 \pm 0.0003$. We have also carried out a series of experiments for testing Glauber's theory of coherence. Given that there exist alternative theories which generalise quantum mechanics beyond second order interference~\cite{zyczkowski2008quartic,dakic2014density,lee2017higher}, testing these theories remains important. There are several possible extensions to this work. On the theoretical side, it would be useful to know which states are optimal in terms of the Fisher information for estimating the probabilities of interest. Experimentally, there are several possible extensions to the work described here, including using squeezed light~\cite{shajilal202212}, performing these tests in alternative reference frames~\cite{fink2017experimental,restuccia2019photon}, and performing alternative tests of quantum mechanics~\cite{mazurek2021experimentally,grabowecky2022experimentally}.

\section{Acknowledgements}
This research was funded by the Australian Research Council Centre of Excellence CE170100012, Laureate Fellowship FL150100019 and the Australian Government Research Training Program Scholarship.

This work was funded by the Deutsche Forschungsgemeinschaft (DFG, German Research Foundation) - Projektnummer 445275953. The authors acknowledge support by the German Space Agency DLR with funds provided by the Federal Ministry for Economic Affairs and Climate Action BMWK under grant number 50WM2165 (QUICK3) and 50RP2200 (QuVeKS). T.V. is funded by the Federal Ministry of Education and Research (BMBF) under grant number 13N16292.

\appendix

\section{Homodyne detection}
\label{apen:homo}
We start with the two states which are sent to the photodetectors in homodyne detection
\begin{equation}
\ket{\alpha_1}=\ket{t\alpha_\text{out}+\sqrt{1-t^2}\text{e}^{\mathrm{i}\phi}\beta}\;,
\end{equation}
and
\begin{equation}
\ket{\alpha_2}=\ket{-\sqrt{1-t^2}\alpha_\text{out}+t\text{e}^{\mathrm{i}\phi}\beta}\;.
\end{equation}
This allows us to write an expression for $\langle\hat{n}\rangle$
\begin{equation}
\begin{split}
\langle \hat{n}_1\rangle=&t^2\abs{\alpha_\text{out}}^2+(1-t^2)\abs{\beta}^2\\&+t\sqrt{1-t^2}\beta(\alpha_\text{out}^*\text{e}^{\mathrm{i}\phi}+\alpha_\text{out}\text{e}^{-\mathrm{i}\phi})\;,
\end{split}
\end{equation}
and 
\begin{equation}
\begin{split}
\langle \hat{n}_2\rangle=&(1-t^2)\abs{\alpha_\text{out}}^2+t^2\abs{\beta}^2\\&-t\sqrt{1-t^2}\beta(\alpha_\text{out}^*\text{e}^{\mathrm{i}\phi}+\alpha_\text{out}\text{e}^{-\mathrm{i}\phi})\;.
\end{split}
\end{equation}
The subtracted photocurrent is proportional to
\begin{equation}
\label{eq:realhomo}
I_1-I_2\propto g_1\langle\hat{n}_1\rangle-g_2\langle\hat{n}_2\rangle\;.
\end{equation}
Setting $g_1=g_2=1$ and $t=1/\sqrt{2}$ gives Eq.~\eqref{eq:homoout}. Using the expression above allows us to model imperfections in homodyne detection.

\section{Quaternion example with $F=0$ }
\label{apen:Feg}
The norm of a quaternion $q=a+\mathrm{i}b+\mathrm{j}c+\mathrm{k}d$ is given by
\begin{equation}
\norm{q}=\sqrt{a^2+b^2+c^2+d^2}\;.
\end{equation}
It will be useful to describe a quaternion as $q=a+v$, with 
\begin{equation}
\norm{v}=\sqrt{b^2+c^2+d^2}\;.
\end{equation}
This allows the exponential of a quaternion to be calculated as
\begin{equation}
\text{e}^q=\text{e}^a\big(\text{cos}(\norm{v})+\frac{v}{\norm{v}}\text{sin}(\norm{v})\big)\;.
\end{equation}

Now consider three coherent states $\ket{\alpha_A}=\ket{\alpha_B}=\ket{\alpha_C}=\ket{1}$ each with a different phase shift, $\phi_{A}$, $\phi_{B}$ and $\phi_{C}$. We allow the phase shift to be a quaternion, $\phi_{A}=\mathrm{i}\pi/2$, $\phi_{B}=\mathrm{j}\pi/2$ and $\phi_{C}=\mathrm{k}\pi/2$. After this phase shift the three coherent states become $\ket{\alpha_A}=\ket{\mathrm{i}}$, $\ket{\alpha_B}=\ket{\mathrm{j}}$ and $\ket{\alpha_C}=\ket{\mathrm{k}}$. With only one coherent state turned on, the three coherent states which reach the detector are scaled by $\sqrt{3}$, \mbox{$\ket{\alpha_{A,\text{out}}}=\ket{\mathrm{i}/\sqrt{3}}$}, $\ket{\alpha_{B,\text{out}}}=\ket{\mathrm{j}/\sqrt{3}}$ and $\ket{\alpha_{C,\text{out}}}=\ket{\mathrm{k}/\sqrt{3}}$. This gives $P_A\propto\norm{\alpha_{A,\text{out}}}^2=1/3$, $P_B\propto1/3$ and $P_C\propto1/3$.

With two coherent states turned on, the three output coherent states are $\ket{\alpha_{A,B,\text{out}}}=\ket{(\alpha_A+\alpha_B)/\sqrt{3}}=\ket{(\mathrm{i}+\mathrm{j})/\sqrt{3}}$, $\ket{\alpha_{A,C,\text{out}}}=\ket{(\mathrm{i}+\mathrm{k})/\sqrt{3}}$ and $\ket{\alpha_{B,C,\text{out}}}=\ket{(\mathrm{j}+\mathrm{k})/\sqrt{3}}$. This gives $P_{A,B}\propto\norm{\alpha_{A,B,\text{out}}}^2=2/3$, $P_{A,C}\propto2/3$ and $P_{B,C}\propto2/3$. In all of these probabilities the constant of proportionality is the same, it is simply normalising by the total mean photon number used, and so can be ignored. These probabilities give $\beta=\gamma=\zeta=0$, which in turn gives $F=0$. This simply serves to illustrate how quaternion quantum mechanics can give different predictions than complex quantum mechanics.

\section{Homodyne detection assuming quaternion coherent states}
\label{apen:homocompl}
In our work we require an additional assumption compared to previous tests of hypercomplex quantum mechanics, highlighted in Section~\ref{sec:expprop}. The assumption is that using measurements proportional to $\abs{\alpha_\text{out}}$, we can infer the photon rate of a given configuration. The key question is whether or not we have already assumed quantum mechanics is complex in our test. In textbook quantum mechanics measuring the $\hat{x}$ quadrature corresponds to measuring the real part of a coherent state amplitude and measuring the $\hat{p}$ quadrature corresponds to measuring the imaginary part. Thus, it appears that complex quantum mechanics is built into a quadrature measurement. However, this is simply because the description of the quadrature operators and coherent states uses conventional quantum mechanics. For example, consider a coherent state $\ket{\alpha}=\ket{\mathrm{i}\alpha_i+\mathrm{j}\alpha_j}$, with $\alpha_i$ and $\alpha_j$ real. (To have eluded experimental discovery so far any prospective $\alpha_j$ terms would likely be small.) According to conventional quantum mechanics, the subtracted photocurrent should satisfy
\begin{equation}
I_1-I_2\propto \text{e}^{\mathrm{i}\phi}\alpha^*+\text{e}^{-\mathrm{i}\phi}\alpha\;.
\end{equation}
For complex $\alpha$ there is no problem in the above equation as $\text{e}^{\mathrm{i}\phi}\alpha^*+\text{e}^{-\mathrm{i}\phi}\alpha$ is a real number. However, assume that $\alpha=\mathrm{i}\alpha_i+\mathrm{j}\alpha_j$ and $\phi=\pi/2$, so that we are measuring the phase quadrature\footnote{We could also imagine extending $\text{e}^{\mathrm{i}\phi}$ to $\text{e}^{\mathrm{i}\phi_\mathrm{i}+\mathrm{j}\phi_\mathrm{j}+\mathrm{k}\phi_\mathrm{k}}$, however, this is unnecessary for our simple example.}. We now find that 
\begin{equation}
I_1-I_2\propto \alpha_i-\mathrm{k}\alpha_j\;.
\end{equation}
which is not a real number. However, any physical measurement can only return real numbers. One possible resolution to this is that $\abs{I_1-I_2}\propto \abs{\alpha_i-\mathrm{k}\alpha_j}$. Note that there are obvious flaws in this model, for instance, we do not propose how to distinguish between $\pm(\alpha_i-\mathrm{k}\alpha_j)$. Nevertheless, in this, extremely simple example, there is a testable difference between complex and quaternion quantum mechanics using our experiment. 

\section{Measurement variance when using displaced squeezed states}
\label{apen:squeezing}
In this Appendix we consider the configuration shown in Fig.~\ref{fig:interf} b), with the input state as a displaced squeezed state and the detection as homodyne detection. To ensure a fair comparison of resources, we will fix the mean photon number of the input state. Given $\langle n_\text{T}\rangle$ photons to use for estimating the probabilities of interest, we can split our photons between squeezing and displacement, so that 
\begin{equation}
\alpha_{\text{in}}=\sqrt{f\langle n_\text{T}\rangle}\;,
\end{equation}
and
\begin{equation}
r_{\text{in}}=\text{arcsinh}[\sqrt{(1-f)\langle n_\text{T}\rangle}]\;,
\end{equation}
where $0\leq f \leq 1$ is the splitting fraction. This ensures the total photon number is always conserved. For a given figure of merit and a fixed input mean photon number, $\langle n_\text{T}\rangle$, we can determine the optimal splitting fraction, $f$.

When the probability of detecting a photon, after normalising to the total mean photon number, is given by $P$, the first two moments of the Gaussian state which reaches the homodyne detector are given by 
\begin{equation}
\mu_{\text{out}}=\begin{pmatrix}
0\\
2\sqrt{P}\alpha_{\text{in}}
\end{pmatrix}\;,
\end{equation}
and
\begin{equation}
\Sigma_{\text{out}}=\begin{pmatrix}
1-P+P\text{e}^{2r_{\text{in}}}&0\\
0&1-P+P\text{e}^{-2r_{\text{in}}}\\
\end{pmatrix}\;.
\end{equation}
We will assume that the measurement strategy is fixed as measuring the $\hat{p}$ quadrature via homodyne detection. Then, given $N$ repetitions of the experiment, the mean and variance of the distribution both depend on the probability which we wish to learn about. 

We first consider the error when estimating the probability from the measured mean. An unbiased estimator for the probability is given by
\begin{equation}
\tilde{P}_\alpha=\bigg(\frac{\mu}{2\alpha_\text{in}}\bigg)^2\;,
\end{equation}
where $\mu$ is the observed mean of the homodyne measurements. The variance in estimating $\mu$ is given by $V_\mu=1-P+Pe^{-2r_{\text{in}}}$. From error propagation formulas, it follows that the variance in estimating $\mu/2\alpha_\text{in}$ is given by $V_\mu/(4\alpha_\text{in}^2)$, and the variance in estimating $P$ is given by $V_{P,\alpha}=V_\mu\mu^2/4$.\footnote{Note that this formula is not valid as $\mu\to0$. For small $\mu$ Eq.~\eqref{Eq:BRprobdistsquared} can be used to infer the variance in estimating the probability.}

Using the measured variance we can also estimate the probability. In this case an unbiased estimator is given by
\begin{equation}
\tilde{P}_r=\frac{1-V}{1-\text{e}^{-2r_{\text{in}}}}\;.
\end{equation}
From the expected moments of a Gaussian distribution the variance in estimating the variance of our distribution is $V_V=2V^2$. The error in estimating $P$ is then given by $V_{P,r}=2V^2/(1-\text{e}^{-2r_{\text{in}}})^2$.

\begin{figure}[t]
\includegraphics[width=0.5\textwidth]{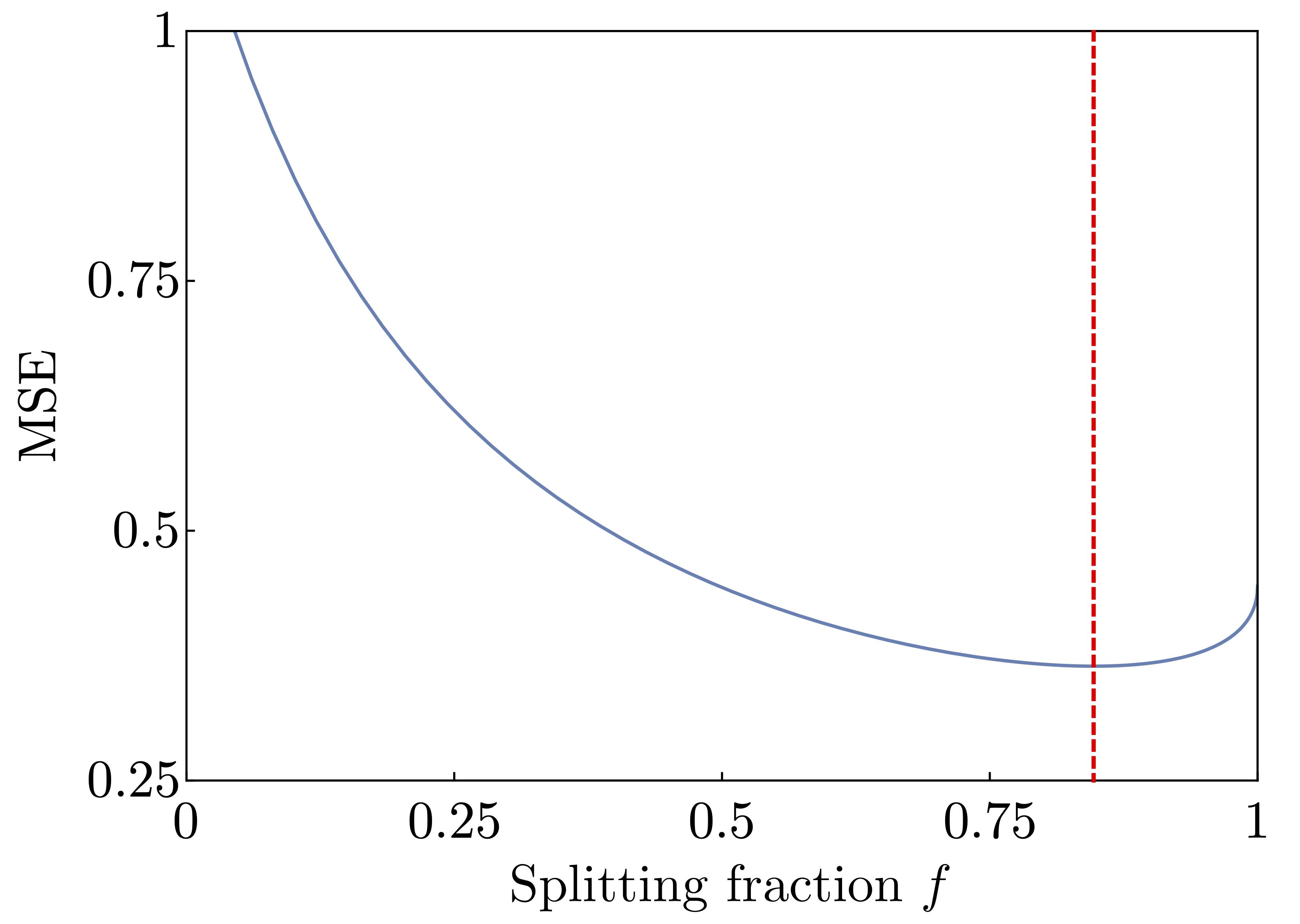}
\caption{\textbf{Mean squared error in estimating the probability $P$ using a displaced squeezed state.} This figure uses $\langle n_\text{T}\rangle=1$ and $P=4/9$, which are experimentally relevant parameters. $f=1$ corresponds to a coherent state. The dashed line corresponds to the minimum MSE value.}
\label{fig:apen:cfi}
\end{figure}

From the homodyne measurement data we therefore obtain two independent estimates of $P$. We can obtain an optimal estimate of $P$ by combining both of these estimates as
\begin{equation}
\tilde{P}=W\tilde{P}_\alpha+(1-W)\tilde{P}_r\;,
\end{equation}
where $0\leq W\leq1$. The error in this estimate is given by
\begin{equation}
V_P=W^2V_{P,\alpha}+(1-W)^2V_{P,r}\;,
\end{equation}
which is minimised when~\cite{li2023optimal}
\begin{equation}
W=\frac{V_{P,r}}{V_{P,\alpha}+V_{P,r}}\;.
\end{equation}
This allows us to determine the optimal splitting fraction, $f$ for a given probability and mean photon number. As shown in Fig.~\ref{fig:apen:cfi}, by allocating some photons to squeezing we can obtain a mean squared error (MSE) which is smaller than that of a coherent state ($f=1$). In this specific example there is a benefit of using non-classical resources like squeezed light. Other quantum resources such as entangled states may provide enhancements in this metrology task~\cite{kacprowicz2010experimental,guo2020distributed}. We can also imagine a scheme where both $\hat{x}$ and $\hat{p}$ quadratures are measured simultaneously. In this case, we have a multiparameter estimation problem, so that collective measurements may prove important~\cite{conlon2023approaching,conlon2021efficient,conlon2022gap}.

\bibliography{HCQMbib}
\end{document}